\documentclass[12pt]{JHEP3}
\usepackage{amsfonts}
\usepackage{amssymb} \usepackage{cite}
\def\be{\begin{equation}}
\def\ee{\end{equation}} \def\bea{\begin{eqnarray}}
\def\eea{\end{eqnarray}} \def\ba{\begin{array}}
\def\ea{\end{array}} \def\ben{\begin{enumerate}}
\def\een{\end{enumerate}} 
  \def\lll{\label}

 \title{  Note on (D6,D8) Bound State, Massive Duality and
Non-commutativity
\thanks{This work is in part supported  by:
AvH -- the Alexander von Humboldt foundation.}}

\author{
 Harvendra Singh\\
 Harish-Chandra Research Institute,
\\ Chhatnag Rd., Jhunsi, Allahabad 211 019, India}

\abstract{
In this paper we study half-supersymmetric (D6,D8) bound state
brane configuration
of  {\it massive} type IIA supergravity. We
show this bound state can also be generated
by using massive T-duality rules of type II strings in $D=9$ and starting
from D7-branes. We write down
corresponding Killing spinors and find that these backgrounds
indeed preserve $16$ supersymmetries like any other D$p$-brane
bound state with $B_{\mu\nu}$ field. We also make a point  on the
{\it massive} nature of $B_{\mu\nu} $ field in this background.
The
Seiberg-Witten limits to obtain
non-commutative Yang-Mills theories in $D=9$  are also
discussed, but the full understanding of such gauge theories
remains unanswered.
}

\preprint{hep-th/0212103}

\begin{document}
\newcommand{\eqn}[1]{(\ref{#1})}
\def\cC{{\cal C}}
\def\cG{{\cal G}}
\def\cd{{\cal D}}
\def\com{{\bf [Clarify it!]}}
\def\a{\alpha}
\def\b{\beta}
\def\g{\gamma}\def\G{\Gamma}
\def\d{\delta}\def\D{\Delta}
\def\ep{\epsilon}
\def\e{\eta}
\def\z{\zeta}
\def\t{\theta}\def\T{\Theta}
\def\l{\lambda}\def\L{\Lambda}
\def\m{\mu}
\def\f{\phi}\def\F{\Phi}
\def\n{\nu}
\def\p{\psi}\def\P{\Psi}
\def\r{\rho}
\def\s{\sigma}\def\S{\Sigma}
\def\x{\chi}
\def\o{\omega}\def\O{\Omega}
\def\k{\kappa}
\def\pa {\partial}
\def\ov{\over}
\def\br{\nonumber\\}
\def\ud{\underline}

\section{Introduction}

The idea of non-commutativity in string theory \cite{witten} has
led to
new insights into the understanding of AdS/CFT conjecture
\cite{maldacena}. It  has
been understood that in the presence of
  NS-NS $B_{\m\n}$ field open strings behave in
such a way that low energy field theory on the anti-de Sitter
(AdS) boundary could be
described by a non-commutative super-Yang-Mills
(NCYM) theory.
The Seiberg-Witten map \cite{witten}
\be\label{sw1}
 G^{\m\n}+ {\t^{\m\n}\ov 2\pi\a'} = {1\ov
g_{\m\n}+2\pi\a'
B_{\m\n}}
\ee
defines for us open string metric, $G_{\m\n}$, and noncommutativity
parameter,
$\t^{\m\n}$, in terms of closed string metric, $g_{\m\n}$, and
background, $B$,
field. The open string coupling is given by
\be\lll{sw2}G_o^2=g_s^2 ~{{\rm det
}(g+B)\over{\rm det} (g)}.
\ee
The Yang-Mills coupling is defined through $g_{YM}^2=G_o$.

Note that the above map works so long as
the closed string backgrounds are constant which is the case for
all D$p$-branes with $2\le p\le6$.
For $p>6$ branes the backgrounds are not asymptotically constant
or flat and the holographic picture is less clear. We
focus in this paper on  D8-branes which have  only  one
transverse
direction.  We would like to study the  dual
Yang-Mills theories for $N$ D8-branes, with or without $B$ field.
It is
not  clear if the above Seiberg-Witten relations hold good when
the supergravity backgrounds are not constant as is the case with
D8-branes. Previously,
$D0-D8$ system with a $B$-field have  been studied in \cite{witt}, see
also \cite{park}.
Interestingly, in a recent paper \cite{sethi},  Hashimoto and Sethi
have studied
holography for  time-dependent backgrounds  assuming
that backgrounds are sufficiently locally constant (see
also \cite{clo}). Results have
been interesting and the application of above Seiberg-Witten
relations reproduces the desired results.

We will employ the similar idea for the D8-branes here and we
assume that
the string backgrounds, though not constant, but are locally constant
so that Seiberg-Witten maps could be applicable.  Under this assumption we
basically study the decoupling limits
of the supergravity backgrounds and  make some observations
about the maximally supersymmetric  NCYM
theories on the nine-dimensional boundaries of AdS$_{10}$ regions.
We find that  under the
decoupling limits  the closed strings indeed get decoupled.

The paper is organized in the following way. In  section-2 we
aim to reconstruct a (D6,D8) bound state with $B$-field by using
massive duality relations in nine dimensions \cite{berg}. We
also obtain the Killing spinors and also discuss the nontrivial
massive nature
of the $B$-field in this background. In  section-3
we study the decoupling limits and discuss the nature of boundary
conformal field theories (CFT) in nine spacetime dimensions. We also
construct
(D4,D6,D8) bound state in section-4. The
conclusions are given in section-5.

 \section{The (D6,D8) bound state}

 Recently, the following background configuration was
obtained in \cite{singh1} as a solution of massive type IIA
supergravity theory \cite{roma}
\bea\label{news1}
&&d  s^2_{10}= H^{1\ov2}\left\{ H^{-1} (-dt^2 +
dx_1^2+\cdots+dx_6^2)  + H'^{-1}( dy_1^2+dy_2^2 )+dz^2\right\}\ ,
\br &&e^{2\f}=g_s^2 H^{-{3\ov2}} H'^{-{1}}\ ,\qquad
dA_{(1)}=-{m~sin\t\ov{g_s}}
dy_1\wedge dy_2\ ,\br
&&  B_{{y_1}{y_2}}= tan\t (1- H'^{-1})
\eea
where  $H=1  + m |z|$
and $H'=1+cos^2\t(H-1)$. Here $m=m_0 g_s /cos\t$ and we
reserve  $m_0$
to denote the mass parameter (cosmological constant) of
the massive type IIA supergravity \cite{roma}. The parameter
 $g_s$  represents string coupling. This
configuration has  nontrivial
$B$-field along with a constant flux of gauge fields and is
interpreted as a bound state of
D6 and D8 branes.  Note
the  value of $ B$ is
$tan\t (1-H'^{-1})$ which  is  usually the case
with all (D$(p-2),$D$p$) bound states for $(2\le p\le 6) $, see
\cite{malrus,roy,lurs}.  It is therefore
interesting to study the non-commutative Yang-Mills
 decoupling limits
\cite{witten} for (D6,D8) bound state \eqn{news1}, which
we will do in the next section.

The bound state \eqn{news1} was obtained by exploiting  the
massive T-duality
symmetries in $D=8$, see  \cite{singh1} for details. However,
we shall  show next that above bound state can also be constructed
by using massive T-duality rules in nine dimensions \cite{berg}.
These nine-dimensional massive T-duality rules were constructed in
order to relate
massive type IIA backgrounds with type IIB backgrounds in nine
dimensions.

\subsection{D7-brane and massive duality in $D=9$}
In the case of asymptotically
flat branes,  in order to construct
(D$(p-2)$,D$p$) bound states with nontrivial $B_{\m\n}$ field
there is
a well known procedure described in \cite{bmm}.\footnote{ D$p$/D$(p-2)$
brane bound states have also been worked out in \cite{costa}.} According
to this method, we
need to start with parallel D$(p-1)$-branes delocalised along one
transverse direction,
$y$
({\it say}), and subsequently make a rotation in a plane involving
the
isometry direction $y$  and a spatial direction parallel to
the branes. A subsequent
application of T-duality along one of the rotated
coordinates generates a solution with $B$-field.
We shall be adopting  this method to obtain
 (D6,D8) brane bound
state.

We start with the delocalised D7-branes in type IIB string theory
given
in \cite{berg},
\bea\label{news1a}
&&d  s^2= H^{1\ov2}\left\{ H^{-1} (-dt^2 +
dx_1^2+\cdots+dx_7^2 )+dz^2+dy^2\right\}\ ,
\br &&e^{2\f^{(b)}}= H^{-2} \ ,\qquad
\x_{(0)}= m~ y \ ,
\eea
with the harmonic function $H(z)=1+m |z|$. (We have set
$g_s=1$ in this section). Note that the
harmonic function is linear in $z$, like in a domain-wall,
and is continuous at $z=0$ where the brane is
localized. However, $\pa_z H$ is discontinuous at that
point. This discontinuity is related to the tension
of delocalised  D7-branes (or D8-branes after duality). Following
\cite{bmm} our
next step would be to make the rotation in $(y,~x_7)$ plane,
\bea
\left(\begin{array}{c}
 y \\ x_7 \end{array}\right)=
\left(\begin{array}{cc}
 cos\t & -sin\t \\ sin\t & cos\t \end{array}\right)
\left(\begin{array}{c}
 y_1 \\ y_2 \end{array}\right)
\eea
In these new coordinates
the solution \eqn{news1a} becomes
\bea\label{news1b}
&&d  s^2= H^{1\ov2}\big[ H^{-1} (-dt^2 +
\sum_{i=1}^6dx_i^2 )+ ({sin^2\t\ov H}
+cos^2\t) dy_1^2\br &&~~~~~~+({cos^2\t\ov H}
+sin^2\t) dy_2^2+ sin2\t (H^{-1}-1)dy_1dy_2+dz^2 \big]\ ,
\br &&e^{2\f^{(b)}}= H^{-{2}} \ ,\qquad
\x_{(0)}= m~ (cos\t y_1 -sin\t y_2)\ .
\eea
Now we would like to make T-duality along $y_1$ direction. One
will note that
neither $y_1$ nor $y_2$ is a isometry direction in the usual
sense because the type IIB axion $\x$ depends linearly on both of
them. However, since the field strength $d\x$ is constant we can
make use of generalized (massive) T-duality rules  constructed
in \cite{berg}. Let us identify the direction $y_1$ to be along
the
circle. This will fix the mass $m_0$ of massive type IIA theory
to be
given by $m\, cos\t$. Using the duality relations in
\cite{berg}
\bea
&& e^{2\f^{(a)}}=e^{2\f^{(b)}}/g^{(b)}_{y_1y_1} \
,~~~~g^{(a)}_{y_1y_1}=1/g^{(b)}_{y_1y_1}
\br && B^{(a)}_{y_1\m}=-g^{(b)}_{y_1\m}/g^{(b)}_{y_1y_1}
, \qquad A_{y_1}= -\x +m_0 y_1
\eea
we obtain  correspondingly a
massive type IIA background
\bea\label{aa}
&&d  s^2_{10}= H^{1\ov2}\left\{ H^{-1} (-dt^2 +
dx_i^2)  + H'^{-1}( dy_1^2+dy_2^2 )+dz^2\right\}\ ,
\br &&e^{2\f^{(a)}}= H^{-{3\ov2}} H'^{-{1}}\ ,\qquad
d A_{(1)}=-m~sin\t dy_1\wedge dy_2\ ,\br
&&  B_{{y_1}{y_2}}^{(a)}= tan\t (1- H'^{-1}) ,
\eea
with  $H=1 + m  |z|$
and $H'=1+cos^2\t~(H-1)$. This is precisely the configuration
written in eq.\eqn{news1}.
Thus we have shown that in two different ways, one as we employed
in \cite{singh1} and the second which we have described in this
section, we lead
to the same end result. This is nothing but  proves the compatibility
of the
 Scherk-Schwarz reductions of massive type IIA
supergravity on $T^1$ \cite{berg} and on
$T^2$ \cite{singh1} with
constant background RR-fluxes.

\subsection{Supersymmetry}

It is presumed that massive T-duality preserves the
supersymmetries of the background configurations in the same way
as the
ordinary T-duality does. Based on this hypothesis we did claim in
\cite{singh1} that (D6,D8) solution preserves 16 supersymmetries
since it had been obtained through an SL(2,R) rotation of the $D8$-brane
solution \cite{singh1}. Let us clarify on the aspects of supersymmetry,
we know that massive
type IIA does not have any maximally supersymmetric ground
state instead the theory admits D8-branes which are half
supersymmetric. On the other hand type IIB supergravity does admit
maximally supersymmetric
Minkowskian ground state and also ${1\ov2}$-supersymmetric
brane configurations including the D7-branes above. Under the $T^1$
compactification these $1/2$-susy
backgrounds are mapped from  IIB side to the massive IIA side
and vice versa \cite{berg}.  Note  that supersymmetries do match
on the both
sides. Thus from this argument also  (D6,D8) bound state
 obtained from D7-branes in last subsection must
have $1/2$ supersymmetries. So we would like to make an
explicit check of the supersymmetries of the (D6,D8)
background in
question and provide explicit solution for the Killing spinors.

Let us first write down most general $SL(2,R)$ covariant
set of (D6,D8) solutions as
\bea\label{ne1}
&&d  s^2_{10}= H^{1\ov2}\left\{ H^{-1} (-dt^2 +
dx_i^2)  + H'^{-1}( dy_1^2+dy_2^2 )+dz^2\right\}\ ,
\br &&e^{2\f}= H^{-{3\ov2}} H'^{-{1}}\ ,~~~
dA_{(1)}=- b~m~
dy_1\wedge dy_2\ ,~~~
  B_{{y_1}{y_2}}= {1\ov d} (b+c H'^{-1})
\eea
where  the harmonic functions are given by
\bea
&&H=1 + m  |z|,~~~ H'=c^2+d^2 H,~~~ m=m_0 /d
\eea
and $m_0$
is the mass parameter of
the massive type IIA supergravity. The real parameters $a,b,c,d$
describe an $SL(2,R)$ matrix $\left(\ba{cc} a&b\\ c&d\ea\right)$.
For the particular choice $\left(\ba{cc} 1&b\\
0&1\ea\right)$ the solution \eqn{ne1} reduces to the D8-brane
and for the case $\left(\ba{cc}  cos\t& sin\t\\
-sin\t&cos\t\ea\right)$  it reduces to the background in
\eqn{news1}.

The supersymmetric variations of dilatino and gravitino can be
obtained from \cite{roma} which for our case are (in
Einstein metric)
\bea
&&\d \l \equiv -{1\ov 2\sqrt{2}}\pa_\m\f \G^\m\ep  -{5\ov
8\sqrt{2}}m_0e^{{5\ov4}\f} \ep +{3\ov
16\sqrt{2}}e^{{3\ov4}\f} F_{\m\n}\G^{\m\n}\G_{11}\ep
+{1\ov
24\sqrt{2}}e^{-{1\ov2}\f} H_{\m\n\l}\G^{\m\n\l}\G_{11}\ep
\br &&
\d \p_\m\equiv D_\m\ep -{1\ov
32}m_0e^{{5\ov4}\f} \G_\m\ep -{1\ov
64}e^{{3\ov4}\f} F_{\n\l}(\G_\m^{~\n\l}-14\d_\m^\n\G^\l)\G_{11}\ep
\br &&~~~~~+{1\ov
48}e^{-{1\ov2}\f}
H_{\r\n\l}(\G_\m^{\r\n\l}-9\d_\m^\r\G^{\n\l})\G_{11}\ep
\eea
where $ F_{(2)}=dA_{(1)} + m_0 B_{(2)}$ and $ H_{(3)}=dB$ and
$\G_{11}$ is the
chirality operator in ten dimensions.
The Killing spinors are those solutions for which these variations
vanish.  The dilatino equation $\d \l=0$  for the
background \eqn{ne1}
simplifies to
\be\lll{ne2}
\big[(c\g_z\g_{y_1}\g_{y_2}\G_{11} +(H')^{1\ov2} ) -d~
H^{1\ov2}\g_z\big]\ep=0
\ee
where all small $\g$ matrices are constant 10-dimensional gamma
matrices and their indices are raised and lowered with the
tangent space metric.  To find a solution of \eqn{ne2} let us make
an ansatz
\be\lll{ne3}
\ep= f \ep_0^{+}+g \ep_0^{-}
\ee
where $\ep_0^{\pm}={1\pm\bar\g\ov2}\ep_0$ with $\bar\g=
\g_z\g_{y_1}\g_{y_2}\G_{11}$. The constant spinors $\ep_0$
satisfy the
condition $\g_z\ep_0=\ep_0$. Note that this projects out 16
spinors  out of  32 constant spinors and  thus  eventually
breaks half of supersymmetries. Substituting the ansatz
\eqn{ne3} in
\eqn{ne2} gives us the following relation between $f$ and $g$,
\be
f={d \sqrt{H}\ov c+\sqrt{H'}} g
\ee
in terms of which $\ep=g \left( {d \sqrt{H}\ov c+\sqrt{H'}}
\ep_0^{+}+\ep_0^{-}\right)$.
The over all function $g$ can be determined by gravitino
variations. Consider the equation $\d \p_z=0$, this implies
that $g$ must satisfy
\be\label{ne4}
\pa_zg-{m\ov 32} ({1\ov H}+{6 c^2\ov HH'}-{8c\ov H\sqrt{H'}})g=0
\ee

Now taking $g$ to be of the form $g\equiv H^p H'^{q} (c +\sqrt{H'})^r$
and substituting it in
the eq. \eqn{ne4} we find that $p,q, r$ have a unique
solution
$$p=-{1\ov32},~~q=-{3\ov16},~~r={1\ov2} .$$

Thus the Killing spinors for the (D6,D8)
background are
\bea\lll{ne5}
\ep&=&H^{-{1\ov
32}}H'^{1\ov16}\left[\left(1-{c\ov\sqrt{H'}}\right)^{1\ov2}
\ep_0^{+}+
\left(1+{c\ov\sqrt{H'}}\right)^{1\ov2}\ep_0^{-}\right].
\eea
It could be checked that all other Killing equations are satisfied
by this solution. When $c=0$  eq.\eqn{ne1}
becomes a  D8-brane background and eq.\eqn{ne5} also reduces to
standard
Killing spinors for these branes. In summary, we have
proved that the
(D6,D8) bound state preserves 16 supercharges  same as
D$8$-branes. Thus the action of massive duality rotations on the
backgrounds do
not break supersymmetries of the backgrounds. It also
indirectly means that massive type II theories obtained through
generalized Scherk-Schwarz reduction of type II supergravities
compactified on $T^2$
are  maximal supergravity theories.

\subsection{Massive $B$-field}

Let us briefly discuss the massive nature of the 2-rank tensor
field $B$ in the background \eqn{news1}. As it can be seen from
the 2-form field strength $ F_{(2)}=dA + m_0 B$ and
the 3-form field strength $H_{(3)}=dB$  that there is a
(stueckelberg) gauge
invariance through which $B$-field can eat one-form $A$ and become
massive. Let us define $B'= B +{1\ov m_0} dA$ and replace
every where in the action
$F_{(2)}\equiv m_0 B'$ while $H=dB'$. Under this gauge fixing
background
\eqn{news1} can be reexpressed as
\bea\label{new11}
&&d  s^2_{10}= H^{1/2}\left\{ H^{-1} (-dt^2 +
dx_i^2)  + H'^{-1}( dy_1^2+dy_2^2 )+dz^2\right\}\ ,
\br &&e^{2\f}=g_s^2 H^{-{3\ov2}} H'^{-{1}}\ ,\qquad
B'=-{tan\t\ov H'}dy_1\wedge dy_2\ ,
\eea
with $ H=1+ m |z|,~H'=1+cos^2\t(H-1)$ and $m=m_0 g_s/
cos\t$. Here $B'$ field is
explicitly massive with mass being $m_0$.\footnote{
Although it is difficult to define a mass in the
domain-wall (curved) backgrounds. Here {\it mass} means that
field has $(mass)^2$ term in the action, see the Appendix.}
Nevertheless background in \eqn{new11} is half-supersymmetric.

The scalar curvature for above background metric is
\be\lll{curvscal}
R=-{14 m'^2 +2mm'(5+19m'|z|)+m^2(21+52m'|z|+45m'^2|z|^2)\ov 4
H^{5\ov2}H'^2}
\ee
where we have  defined $m'=m cos^2\t$.
This result will be used in the next
section. When $\t=0$, $m'$ becomes equal to $m$ and the expression in
eq.\eqn{curvscal} reduces to the curvature for pure D8-brane
background.

So far we chose to keep plus sign in the harmonic function of the type
$H=1 \pm m |z|$ although
solutions exist with both the signs.  It has
been found in \cite{polwit} that for D8-branes with $+$ve tension
\footnote{Tension
of D$p$-brane is defined as $T_p\sim{1\ov g_s (\a')^{p+1\ov2}}$},
a lower sign in the
harmonic function $H=1\pm m|z|$ is favored. If we use a
negative sign in the harmonic function $H$ in \eqn{new11},
it follows
that as we go far away from the
8-branes not only string coupling but also $B$-field
diverges. Thus such D8-brane
configurations cannot be defined independently and far away
from
the orientifold 8-planes \cite{polwit}. Near the O8-planes the
above geometry has to
take into account the back reaction from the orientifolds
also and the geometry
will be appropriately modified. In terms of string
quantities $g_s $ and $\a'$, the type IIA mass
parameter $m_0$ can  suitably
be expressed as $m_0= {c_8 N \ov \sqrt{\a'}}$, with $c_8$ being
an appropriate combinatoric factor and $N$  the number of
D8-branes.

\section{Non-commutative Field Theories}

 We are now ready to formulate a discussion in the field theory
direction. It is well known fact that in the Seiberg-Witten limit
($\a'\to0$) the closed string backgrounds describe holographic
dual picture of the
boundary conformal field theories (CFTs) in various brane
pictures. Precisely,  a CFT defined on the boundary
of an anti-de Sitter (AdS) spacetime is holographic dual to the
gravity (string) theory in the bulk
which  constitutes the  AdS space \cite{maldacena}. Near
horizon limits
 of various brane solutions in (M-)string theories
give rise to AdS spacetimes. We would like to see whether the same
picture of AdS/CFT  emerges in the case of D8-branes
also. Since
D8-branes are not asymptotically flat we have to be careful.
\subsection{No $B$ field}
We  consider O8-D8 combination  so
we are eventually in type I$'$ picture \cite{polwit}. We will
shall first consider the case without $B$-field. Let us
consider $N~(N<8) $ positive tension D8-branes
situated at one of the orientifold plane. Including the
backreaction of the O8-plane the background geometry
for $N$ D8-branes
can be written as ($i.e.$ with an effective mass parameter $m_0= {c_8
(8-N)\ov\sqrt{\a'}}$)
\bea\label{new1}
&&d  s^2_{10}= H^{1\ov2}\left\{ H^{-1} (-dt^2 +
dx_1^2+\cdots+dx_8^2  )+dz^2\right\}\ ,
\br &&e^{2\f}=g_s^2 H^{-{5\ov2}}\ ,\qquad
 H=1+ {c_8 (8-N)g_s\ov\sqrt{\a'}} |z|\ .
\eea
Thus so long as $N<8$ we can consider the following decoupling
limit,
in analogy with other D$p$-branes \cite{itzhaki},
\bea
&&\a'\to 0 ,\qquad  |z|\to\a' u,\qquad g_s\to \tilde g
(\a')^{-{5\ov2}},\qquad
g_{YM}^2\tilde N=fixed,\br
\eea
where various parameters $\tilde g(=g_{YM}^2),~ \tilde N$ and the energy
scale $u$ (the expectation value of the Higgs) are kept
fixed. We
are using notation $\tilde N=c_8(8-N)$ in order to distinguish it
from
$N$, the number of parallel D8-branes. Note that  D8
background is not asymptotically flat nevertheless we shall
implement above scaling limit. Under
this limit $H\sim \tilde g \tilde Nu/\a'^2$ and \eqn{new1}
becomes \bea\label{new1a}
&&ds^2\sim \a' \sqrt{\tilde g \tilde Nu^5}\big[ {1\ov \tilde g
 \tilde Nu^3}
(-dt^2
+dx_1^2+\cdots+dx_8^2)+ {du^2\ov u^2}\big],\br
 && e^{2\f}\sim {1\ov \tilde N^2}{1\ov g_{eff}}
\eea
where effective super-Yang-Mills coupling at the scale $u$ is
defined as
$g_{eff}^2=\tilde g \tilde N u^5$. It can be seen that the expression
within angular brackets on the
r.h.s. of \eqn{new1a} is a space-time filling  AdS$_{10}$
geometry.\footnote{The decoupled background in \eqn{new1a},
which is conformally $AdS_{10}$ type,
is  nevertheless a solution of massive type IIA supergravity with
an effective mass
parameter
$m_0=\tilde N / \sqrt{\a'}$, for any  value of $\a'$. }
Therefore we
can discuss holographic
field theory on the nine-dimensional boundary of AdS spacetime in
this decoupling limit. The background does not have
transverse isometries, so the boundary field theory in nine
dimensions would
be ${\cal N}=1$  super-Yang-Mills theory with a gauge group $SO(2N)$
for all $N<8$.
 There is no R-symmetry in the gauge theory because the D8
background has only one transverse direction.
 From eq.\eqn{curvscal} we find that the
 curvature scalar  in string
units is given by
$$-\a' R={21\ov 4} \sqrt{1\ov {\tilde g  \tilde Nu^5}}={21\ov4}
{1\ov g_{eff}}.$$
Thus in the IR region,
$u^5\ll{1\ov \tilde g\tilde N}$,  where $g_{eff}^2\ll1$, super
Yang-Mills
description holds good. But in this region the curvature and
string coupling are both
large and the supergravity
is not a valid description. While in the UV region, $u^5\gg{1\ov
\tilde g \tilde N}$, curvature and string coupling are small and
low energy
sugra is a valid description. It is useful since in UV
region $g_{eff}\gg1$
and the field theory breaks down at some point. Since field
theories in $D>4$
show bad UV behaviour, it is  useful that supergravity can make
sense out there. However the ten-dimensional Newton's constant
$G_N^{10}\sim g_s^2 \a'^4 $ goes as $1/\a'$ in the decoupling
limit and thus blows up. Which is some what contrary to what one expects
in the decoupling limits. We will see next that Newton's constant indeed
vanishes
if $B$-field is present.
\subsection{$B\ne 0$}
Now we go over to the case of D8-branes where $B$-field is present. The
background in discussion here is given in eq.\eqn{new11}. Again
the  background is not asymptotically flat but we
will insist that
the background variations are small enough locally so that we
can implement the
decoupling limit.
In this case the decoupling limits are slightly modified as
\bea\label{dec1a}
&&\a'\to 0 ,\qquad g_s\to \tilde g (\a')^{-3\ov2},\qquad
g_{YM}^2\tilde N=fixed,\br
&& |z|\to\a' u,\qquad cos\t\to {\a'\ov b},\qquad
(y_1,y_2)\to {\a'\ov b}(\tilde y_1,\tilde y_2).
\eea
Various parameters $\tilde g,~u,~b,~ \tilde N$ are held
fixed and $b$ is the noncommutativity parameter. Under these
limits the harmonic functions
in  \eqn{new11} become (with $m_0={c_8(8-N)\ov\sqrt{\a'}}$)
\be
H\sim {\tilde g \tilde N b u\ov\a'^2},\qquad H'=1+{\tilde g\tilde N u\ov
b}
\ee
and
\bea\label{new1b}
&&ds^2\sim \a' \sqrt{\tilde gb\tilde N u^5}\bigg[ {1\ov \tilde g
b\tilde N u^3}
\left(-dt^2
+\sum_{i=1}^6dx_i^2+(1+{b\ov \tilde g \tilde Nu})^{-1}(d\tilde
y_1^2+d\tilde y_2^2)\right)+ {du^2\ov
u^2}\bigg],\br
 && e^{2\f}\sim {\tilde g^2 \ov  (\tilde g \tilde N
b u)^{3\ov2}}{1\ov H'} \ ,\qquad
B'={\a'\ov b}(1+{ \tilde g \tilde Nu\ov b})^{-1}d\tilde y_1d\tilde
y_2
\eea
where effective gauge coupling of nine-dimensional NCYM at the
scale $u$ is defined as
$g_{eff}^2=\tilde g b\tilde Nu^5$.
Note that the ten-dimensional Newton's constant $G^{10}_N$
is of the order of $\a'$ and  vanishes in the limit
$\a'\to 0$.  This is a sign that closed strings indeed get decoupled in
the $\a'\to0$ limit when $B$-field is present.
Note from \eqn{new1b} that after the decoupling
limit {\it massive}
$B'$ field precisely behaves as in 4-dimensional NCYM theories
\cite{malrus,roy,lurs}.

The  decoupling limits
\eqn{dec1a} and the
decoupled geometry \eqn{new1b} do indicate that there is a dual NCYM
theory, but where
does this NCYM live? In the type-I' theory there are two orientifold
fixed planes, one at $z=0$ and other at $z=\pi$, and 16 D8-branes are
sandwiched between
these two fixed points. At the
fixed points the NS-NS $B$-field vanishes, as it can  be  seen from
eq.\eqn{news1} also.
Therefore, there cannot be any noncommutativity if we place $N$ D8-branes
right at the fixed
point $z=0$ and rest $(16-N)$ at the other fixed point. However if
we place $N$ D8-branes at some finite distance away, say
at $z=z_0$, there is a
nonvanishing $B$-field background there.\footnote{ The harmonic
function $H(z)$ in \eqn{news1} and \eqn{new11} would
become $H= 1+{c_8 8\ov\sqrt{\a'}}z_0 +
{c_8(8-N)\ov\sqrt{\a'}}(z-z_0)$
in the region $z_0<z< \pi$ since cosmological constant $m_0$ jumps between
the branes.} Note that such a $z$-dependent $B$-field along the
world-volume directions of the D-brane is not projected
out under orientifolding in type I' theory and nor the constant RR
2-form 'flux' in \eqn{news1} \cite{sena}.\footnote{I thank Ashoke for
discussion on these aspects.}
Under the scaling limit
\eqn{dec1a}, $z_0\to \a' \l$ and $(z-z_0)\to \a' u$,
where  $\l$ would act like a IR cutoff in the YM theory and it will
measure the separation
between O8-plane and $N$ D8-branes.\footnote{The actual
expression for $g_{eff}^2\sim  \tilde g b(1+{8\l\ov(8-N) u})u^5$.
Effectively speaking $\l$
would become a  UV cutoff for the case when $N>8$.} While dealing with
the decoupling limits \eqn{dec1a}, we have ignored $\l$ assuming that
it is infinitesimal. The resultant background  written
in \eqn{new1b} should be seen from that perspective.
These D8-branes under the
decoupling limit will be described by a non-commutative Yang-Mills with
gauge group  $SU(N)$ with $N<8$.

 Now,  it is easy to see from eq.\eqn{new1b}
 that only when $u\gg{b\ov \tilde g\tilde N}$ $(i.e.~ b^2
u^4\ll g_{eff}^2)$, the
geometry in \eqn{new1b} becomes conformally AdS$_{10}$ and the whole
background reduces to that in \eqn{new1a}. Thus we have
a commutative phase in YM theory in the UV region.
But in the UV region effective
gauge coupling is large, so field theory is not quite well defined.
However,
the dual supergravity description holds good
since string coupling and the curvature
\be\label{fUV}
 e^{2\f_{UV}} \sim {1\ov \tilde N^2}{1\ov g_{eff}},
~~~~-\a'R_{UV}={45\ov4}{1\ov g_{eff}} \ee
both are small in UV region.

 Let us go to the IR region where $u\ll{b\ov \tilde g\tilde N }$ $(i.e.~
b^2 u^4\gg g_{eff}^2)$. We have
a non-commutative
phase where the coordinates $\tilde y_1,\tilde y_2$ are
noncommutative; $i.e.$ $[\tilde y_1, \tilde y_2]\sim b$. In IR region
string coupling and
 curvature are given by
\be\lll{fIR}
 e^{2\f_{IR}} \sim {1\ov \tilde N^2}{ g_{eff}\ov b^2 u^4},
~~~~-\a'R_{IR}={21\ov4}{1\ov g_{eff}}. \ee
In the IR region where $b^2 u^4\gg g_{eff}^2\gg 1~( i.e. {1\ov ( \tilde
g\tilde N b)^{1\ov5}} \ll u \ll {b\ov \tilde
g\tilde N} )$ the
string coupling and curvature are small and sugra description
holds good. This region can be approached if parameters are
chosen such that
$b^{3\ov2}\ggg \tilde g \tilde N$. Since $g_{eff}^2\gg 1$
the NCYM  is strongly
coupled. Further towards the lower IR region $u\ll {1\ov ( \tilde
g\tilde N b)^{1\ov5}}$ and into deep IR region,  both the
string quantities are large, but the field theory
description is perturbatively well defined
due to the weak gauge coupling, $ g_{eff}^2\ll 1$.

In the strong string coupling region type IIA brane systems are well
described
only in an appropriate
M-theory picture. Note that (D6,D8) background is a solution of Romans'
theory which has no straight forward M-theory relationship, see
\cite{hull,hls,singh1}. We shall
describe next a possible way to go to M-theory side based on the approach
in \cite{singh1}.

\subsection{(M5,KK) bound state}
The M-theory background can be obtained by mapping (D6,D8) solution first
to (D4,D6) solution of type IIA supergravity in the following way.
We start with (D6,D8) bound state \eqn{news1} and compactify two
coordinates, $x_5,x_6$, on a $T^2$. Then we follow it up with an
SL(2,R) rotation
$\left(\ba{cc}  0& -1\\
1&0\ea\right)$.
Up-lifting the rotated 8-dimensional
configuration back to ten
dimensions (using the rules described in \cite{singh1})
would give us following (D4,D6) configuration of type IIA
\bea\label{news100}
&&d  s^2_{10}= H^{1\ov2}\left\{ H^{-1} (-dt^2
+\sum_{i=1}^4dx_i^2)
+ H'^{-1}( dy_1^2+dy_2^2 )+(dz^2+dx_5^2+dx_6^2)\right\}\ ,
\br &&e^{2\f}=g_s^2 H^{-{1\ov2}} H'^{-{1}}\ ,\qquad
dC_{(3)}={-m~ sin\t\ov{g_s}}
dy_1\wedge dy_2\wedge~dx_5\wedge~dx_6,\br &&
dA_{(1)}= {m~cos\t \ov{g_s}}
dx_5\wedge dx_6\ ,~~~~~
  B_{(2)}= tan\t (1-H'^{-1})dy_1\wedge dy_2
\eea
with  harmonic functions $H=1  + m |z|,~H'=1+cos^2\t(H-1)$. The parameter
$m$
will be appropriately related to the relevant stringy quantities. This
type IIA
configuration is delocalized (smeared) over transverse $x_5,x_6$
$T^2$-plane.
We can now easily lift this type IIA background to M-theory  solution
\bea\label{M100}
&&d  s^2_{11}= e^{4\f/3}(dx_{11} + A_{(1)})^2
+e^{-2\f/3}ds_{10}^2\br
&&~~~~~=\left({H\ov H'}\right)^{-{1\ov3}}\left\{  (-dt^2
+\sum_{i=1}^4dx_i^2) +(H')^{-1} [dx_{11} + {m\ov2} cos\t (x_5 dx_6 -x_6
dx_5)]^2\right\}\br
&&\hskip2.5cm+ \left({H\ov H'}\right)^{2\ov3}\left\{ dy_1^2+dy_2^2
+H'(dz^2+dx_5^2+dx_6^2)\right\} \ ,
\br &&
G_{(4)}\equiv dC_{(3)}={-m~ sin\t}
dy_1\wedge dy_2\wedge~dx_5\wedge~dx_6+
{m~cos\t sin\t\ov H'^2}
dz\wedge dx_{11}\wedge dy_1\wedge dy_2\ ,\br
\eea
with $H=1  + m |z|,~H'=1+cos^2\t(H-1)$,
where $m$ is now related to the M-theory quantities
as $m\sim { N_5 l_p^3\ov a_x a_y}$.\footnote{ The radius,
$R_{11}$, of the circle coordinate $x_{11}$ is related to the
string coupling as $R_{11}=e^{2\f/3} l_p$ and 11-dimensional Planck length
as $l_p^2=g_s^{2/3} \a'$. $N_5$ is the number of M5-branes , $a_x$ and
$a_y$ are related to the sizes of the two transverse $T^2$s, $x_5,x_6$ and
$y_1,y_2$  respectively.}   This solution represents a bound state system
of M5-brane and Kaluza-Klein (Taub-NUT) monopoles and is smeared over two
transverse $T^2$s. Coordinates $t, x_1,\cdots, x_4, x_{11}$ are along
M5-branes while $C_{\m\n\l}$-field is along $x_{11}, y_1,y_2$ which is
responsible
for having Taub-NUT (TN) charges in this background.  When $\t=0$ in
\eqn{M100} the
background reduces to  $TN \times Mink_7$  \cite{singh1}. If we
set $\t=\pi/2$ then solution reduces to pure M5-branes with $G$-flux over
$T^2\times T^2$.

It should be clear that solution \eqn{M100} represents
an equivalent M-theory background for  (D6,D8)
solution with $B$-field.
It is rather appropriate to discuss decoupling limits of this solution
when string coupling becomes large. Corresponding scaling limits for
\eqn{M100} when $\a'\to0$ can be determined and these are
\bea\label{M101}
&&{|z|\ov \a'}=u=fixed,~~~~l_p\to(\a')^{1/3},~~~R_{11}=N_5=Fixed \br
&&cos\t\to{\a'\ov b},~~~a_x\to\a'^2 \tilde a_x,~~~a_y\to\a'^2 \tilde a_y
\eea
with $\tilde a_x$ and $\tilde a_y$ are fixed area parameters.
Note that the areas of transverse $T^2$s also shrink to zero
under this scaling.
It can be checked
that the background \eqn{M100} indeed gets decoupled in the limit
\eqn{M101}. So in the IR region where the size of eleventh
dimension measured in Planck units $R_{11}(u)/l_p =e^{2\f/3}$ becomes large
it is useful to study above decoupling limits where $l_p\to 0$. The
corresponding boundary field
theory  would be a  non-local $6D$ (0,2) SCFT on a circle
\cite{itzhaki}. The nonlocality arises due to the presence of
Taub-NUT charges in the M5-brane solutions.\footnote{ See
\cite{keshav} for nonlocal 6-dimensional field theories.}
Let us note down the
curvature of the 11-dimensional spacetime measured in the Planck units in
the IR region (using eq. \eqn{fIR})
\be
l_p^2 R\sim e^{2\f/3} (\a' R) \approx  \left({1\ov
\tilde N^2 g_{eff}^2 b^2 u^4}\right)^{1\ov3}.
\ee
The eleven-dimensional curvature measured in Planck units is still large
when $u\to0$.
Therefore this low energy supergravity description will not be reliable as
corresponding (M5,KK) backgrounds would receive higher curvature
corrections. But as
we saw  NCYM and the CFT theories in this region are weakly
coupled and
can make a good description.

\section{(D4,D6,D8) bound state}

It is desirable to obtain D8-branes with $B$-field of higher rank.
To obtain such solutions we can
apply the same method described in section-5 of \cite{singh1}
which led to the construction of (D6,D8) solution. We
start with (D6,D8) bound state \eqn{news1} and compactify two
coordinates, $x_5,x_6$, on $T^2$. Then follow it up with an
SL(2,R) rotation
$\left(\ba{cc}  cos\p& sin\p\\
-sin\p&cos\p\ea\right)$.
Up-lifting the rotated 8-dimensional
configuration  to ten
dimensions (using the rules described in \cite{singh1})
would give us following new configuration of massive type IIA
supergravity,
\bea\label{news101}
&&d  s^2_{10}= H^{1\ov2}\left\{ H^{-1} (-dt^2
+\sum_{i=1}^4dx_i^2
)+f^{-1}(dx_5^2+dx_6^2)
+ H'^{-1}( dy_1^2+dy_2^2 )+dz^2\right\}\ ,
\br &&e^{2\f}=g_s^2 H^{-{1\ov2}}f^{-1} H'^{-{1}}\ ,\qquad
dC_{(3)}={m~sin\p sin\t\ov{g_s}}
dy_1\wedge dy_2\wedge~dx_5\wedge~dx_6,\br &&
dA_{(1)}=-{m~cos\p sin\t\ov{g_s}}
dy_1\wedge dy_2 - {m~cos\t sin\p\ov{g_s}}
dx_5\wedge dx_6\ ,\br
&&  B_{(2)}= tan\t (1-
H'^{-1})dy_1\wedge dy_2+tan\p(1-f^{-1})dx_5\wedge dx_6
\eea
with  harmonic functions $H=1  + m |z|,~H'=1+cos^2\t(H-1)$
and $f=1+cos^2\p(H-1)$. Here parameter $m={m_0 g_s \ov
cos\p~cos\t}$ and as usual
  $m_0$  denotes the mass (cosmological constant) of
the massive type IIA supergravity. This solution has sixteen
supersymmetries and can be described as a bound state of D4, D6
and D8-branes as corresponding magnetic charges are present in
this solution. Note that the $B$-field
in the above solution has rank four while in the (D6,D8) solution
it had rank two only.
One may also describe NCYM decoupling limits for this bound
state as well, similar to the case of (D6,D8) solution, but we
simply do not attempt it here.

\section{Summary}
In this paper we have shown that the (D6,D8) bound state
\cite{singh1} with $B$
field can also be obtained by using T-duality map between
massive-type-IIA supergravity and type-IIB supergravity in $D=9$
\cite{berg}.  We have also explicitly written down the  Killing
spinors which are preserved by this bound state configuration. We
find that though $B$-field is explicitly massive the
(D6,D8) background
preserves 16 supersymmetries.

We have then studied Yang-Mills decoupling limits and have
discussed the behaviour of field theories  at various
energy scales. We are
surprised to note  that these 9-dimensional super-Yang-Mills
theories with maximal supersymmetries are non-commutative in the
IR region while they become
commutative in UV region. This is quite opposite to what we
observe in the case of NCYMs in four dimensions where non-commutativity
appears only in the UV region and it disappears as we go to
IR region and the theories become  ordinary super-Yang-Mills.
On one hand this may
not surprise us so much as we know that noncommutative field
theories any way show UV/IR mixing.

 Thus the appearance of non-commutativity as we go to IR region is
some what very peculiar feature of the nine-dimensional NCYMs
presented here.
We could not understand this unusual behaviour of the
$D=9$ NCYMs, nevertheless we are able to expose this property simply by
studying the decoupling limits involving D8-branes with $B$-field.
From section-3 we note that there is a decreasing jump in the spacetime
curvature  as we move from the IR region
to UV region of dual NCYM theory. Since $g^{UV}_{eff}\gg
g^{IR}_{eff}$, the AdS curvature is
more in the IR region as compared to
the UV region. This would mean the NCYM theory  flows from  higher
curvature (weak gauge coupling) IR
region to a smaller curvature (strong gauge coupling) UV point.
It is not unusual to have such a flow, the gauge theories already in
five dimensions  flow to strong coupling ($g_{YM}=\infty$) UV
fixed point \cite{seiberg} where gauge symmetry enhancement takes place.
There the symmetries are
enhanced to exceptional groups $E_{N+1}$. These gauge groups could be any
$E_8,~E_7,~E_6,~E_5=Spin(10),~E_4=SU(5),~E_3=SU(3)\times
SU(2),   ~E_2=SU(2)\times U(1)$ and $ ~E_1=SU(2)$ depending upon
the number, $N$, of D8-branes present at the orientifold.
Therefore in UV region, the 9D NCYMs must flow to these enhanced symmetry
fixed points
where commutativity is also restored.

Finally, we note that in a recent paper \cite{sethi} it has been
observed, that for
non-constant (but slowly varying) closed string backgrounds,
$g_{\m\n},~B_{\m\n}$, the  Seiberg-Witten
relations give rise to  open string metric $G$ and
noncommutativity parameter $\t$ which are spacetime
dependent. For constant $g$ and $B$ the open string metric and
noncommutativity parameter are however constant as well as the
Yang-Mills coupling. We have not tried to obtain these open string
quantities for D8-branes with $B$ field. In our case
both $g$ and $B$, however, depend on the holographic coordinate $z$
itself.
We do  expect, in general, open string metric and non-commutativity
parameter also
to be  dependent on $z$ ($i.e.$ $u$). Lastly, since $z$ is a coordinate
transverse to the brane directions the
Moyal star product $f*g$  should
be well defined locally (at any given position $z=z_0$ of the boundary).
It will also be associative.
The $z$-dependence is probably an indication  of the fact that
nine-dimensional
NCYMs are nonrenormalizable and heavily cut-off dependent.

\acknowledgments
I would like to thank  D. Ghosal,
R. Gopakumar,  D. Jatkar and A. Sen for  helpful discussions.
This work commenced when I was a member of the theory group at
Fachbereich Physik, Martin-Luther-Universit\"at, Halle
(Germany) for
which I would like to thank Jan Louis for
great hospitality and for providing enriching environment for
research.

\vskip .5cm
\appendix{
\section{\underline {\bf  Romans' type IIA supergravity }}

The 10-dimensional type IIA supergravity, which
describes the low
energy limit of type IIA superstrings,
contains in the massless bosonic spectrum the graviton
$ g_{MN}$, the dilaton
$\f$,  NS-NS two-form $ B_{(2)}$, a R-R one-form $ A_{(1)}$
and a R-R three-form $ C_{(3)}$. The fermionic
sector consists of two gravitini and two Majorana
${1\ov2}$-spinors.
The Romans' supergravity theory  \cite{roma} is a
generalization of the type IIA supergravity to include a mass term
for the NS-NS $ B$-field without
disturbing the supersymmetry content of the theory.
The  bosonic action for Romans' theory in the
string frame can be written  as (after some rescalings)\footnote{
Our conventions are same as in
  \cite{hls}  where
 every product of forms is understood
 to be a wedge product. We denote a $p$-form  with a lower index
like $(p)$.  }
\begin{eqnarray} \label{massive2a}
S&=&\int \bigg[ e^{-2\f}\left\{
R~^{\ast}1+4d\f~^{\ast} d\f -
{1\over2} H_{(3)} ~^{\ast} H_{(3)}\right\}
-{1\over2} F_{(2)} ~^{\ast} F_{(2)} -{1\over2} F_{(4)}
~^{\ast} F_{(4)}- {m_0^2\ov 2 }~^{\ast} 1 \br
&& +{1\ov2} d C_{(3)} d C_{(3)}  B_{(2)} +{1\ov
2}d C_{(3)} d A_{(1)}  B_{(2)}^2+{1\ov3!}d A_{(1)}
d A_{(1)} B_{(2)}^3
+{1\ov3!}m_0 d C_{(3)} B_{(2)}^3 \br
&&+{1\ov8}m_0 d A_{(1)}  B_{(2)}^4+ {1\ov40}m_0^2
B_{(2)}^5\bigg]\ ,
\lll{1aa}
\end{eqnarray}
where $m_0$ is the mass parameter. The
 field strengths in the action \eqn{1aa}
are given by
\be
 H_{(3)}=d B_{(2)}\ ,
\qquad
 F_{(2)}= d A_{(1)} +m_0 B_{(2)}\ ,
\qquad
 F_{(4)}= d C_{(3)}
+ B_{(2)}d A_{(1)}+{m_0\ov2} B_{(2)}^2\ .
\lll{a22}\ee
Note that potentials
$ A$ and $ C$  appear only through their derivatives
in the action \eqn{1aa} and thus obey the standard
$p$-form gauge invariance $A_{(p)} \to A_{(p)} + d\lambda_{(p-1)}$.
The two-form $  B$ on the other hand also appears
without derivatives but nevertheless the `Stueckelberg'
gauge transformation
\be\label{stueck}
\delta  A=-m_0\lambda_{(1)}\ ,\qquad
\delta  B=d\lambda_{(1)}\ ,\qquad
\delta C=-\lambda_{(1)} d A\
\ee
leaves the action invariant.

Now, if we define, $d A + m_0
B=m_0  B',~  C'_3= C_3 -{1\ov 2m_0}  A d A$
then $ H=d B',~ F_{(4)}=d C'+{m_0\ov2}B'B'$. The
above action reduces to
\begin{eqnarray} \label{massive2ab}
S&=&\int \bigg[ e^{-2\f}\left\{
R~^{\ast}1+4d\f~^{\ast} d\f -
{1\over2}  H_{(3)} ~^{\ast} H_{(3)}\right\}
-{1\over2}m_0^2 B' ~^{\ast} B'
-{1\over2} F_{(4)}
~^{\ast} F_{(4)}- {m_0^2\ov 2 }~^{\ast} 1 \br
&& +{1\ov2} d C'_{(3)} d C'_{(3)}  B'_{(2)}
+{1\ov3!}m_0 d C'_{(3)}( B'_{(2)})^3
+ {1\ov40}m_0^2(
B'_{(2)})^5\bigg]\ ,
\lll{1a}
\end{eqnarray}

For the  kind of backgrounds in \eqn{new11} for which
$B'\wedge B'=0,~C'=0$
above action reduces to (with fermionic backgrounds vanishing)
\begin{eqnarray} \label{massive2abc}
S&=&\int \bigg[ e^{-2\f}\left\{
R~^{\ast}1+4d\f~^{\ast} d\f -
{1\over2}  H_{(3)} ~^{\ast} H_{(3)}\right\}
-{1\over2}m_0^2 B' ~^{\ast} B'
- {m_0^2\ov 2 }~^{\ast} 1\bigg]
\eea
which involves  an explicit mass term for $B'$ field and a
cosmological constant term.
}

\end{document}